\documentclass[sigconf]{acmart}


\AtBeginDocument{%
  }

\copyrightyear{2026}
\acmYear{2026}
\setcopyright{cc}
\setcctype{by}
\acmConference[RecSys '26]{20th ACM Conference on Recommender Systems}{September 27-October 02, 2026}{Minneapolis, MN, USA}
\acmBooktitle{20th ACM Conference on Recommender Systems (RecSys '26), September 27-October 02, 2026, Minneapolis, MN, USA}
\acmDOI{10.1145/3773078.3831809}
\acmISBN{979-8-4007-2284-4/2026/09}

\begin{document}

\title{Impression Share Prediction: An Offline Evaluation Task for Ranking Systems}

\author{Mohsen Malmir}
\email{mohsenm@meta.com}
\orcid{0000-0002-3977-1473}
\affiliation{%
  \institution{Meta Platforms, Inc.}
  \country{USA}
}

\author{Houssam Nassif}
\email{houssamn@meta.com}
\orcid{0000-0002-0236-2385}
\affiliation{%
  \institution{Meta Platforms, Inc.}
  \country{USA}
}

\author{Danish Nasir Shaikh}
\email{danishshaikh556@meta.com}
\orcid{0009-0003-5010-2689}
\affiliation{%
  \institution{Meta Platforms, Inc.}
  \country{USA}
}

\author{Taher Rahgooy}
\email{trahgooy@meta.com}
\orcid{0000-0002-2889-0522}
\affiliation{%
  \institution{Meta Platforms, Inc.}
  \country{USA}
}

\author{Murat Ali Bayir}
\email{mbayir@meta.com}
\orcid{0000-0003-0311-3994}
\affiliation{%
  \institution{Meta Platforms, Inc.}
  \country{USA}
}

\begin{abstract}
\par{Offline evaluation is a major gateway before online evaluation of ranking models in A/B testing. Standard offline metrics measure predictive accuracy, but are only a surrogate for downstream utility: a model can improve them while redistributing impressions across objective buckets in ways that degrade downstream utility. No offline method surfaces these impression share shifts before online evaluation. We propose \emph{impression share prediction} as an offline evaluation task: given a candidate ranking model, predict the distribution of impressions it would produce across objective buckets - impressions grouped by optimization goal (e.g., click, video view). The task is inherently counterfactual, since the candidate has never served live traffic. We propose a structural causal model of how model predictions and delivery capacity jointly determine impression allocation, and show the counterfactual effect is identified from observational data. Building on this, we develop a statistical learning framework that predicts impression shares from a candidate's early-interaction confidence signals and current system state, trained on historical data. On data from multiple ranking model families, a Random Forest reduces L1 error by 49\% over a constant baseline for models seen during training. For held-out models, evaluated by time since first appearance, the first hour is the closest analog to true online evaluation and the hardest: the Random Forest falls below the baseline because the capacity state still reflects the prior model. An encoder-conditioned architecture that simulates a 2-hour rollout over recent auction dynamics recovers $+$22\% L1 in this regime.}
\end{abstract}



\maketitle

\section{Introduction}
\par{Offline evaluation is a major gate before A/B testing in ranking systems~\cite{castells2022offline,hidasi2023flaws,gupta2023safe,radwan2024counterfactual}. Impressions are organized by optimization goal - click, video view, and so on - forming a set of objective buckets, each generating distinct downstream utility when an impression leads to its target action. Standard offline metrics measure predictive accuracy on logged data~\cite{li2011replay,swaminathan2015crm_jmlr,dudik2011dr,joachims2017unbiased,swaminathan2017slateope} but are only a surrogate for downstream utility - the total outcome those impressions generate. A model that improves these metrics may still degrade downstream utility~\cite{Li2027optimalPolicies}: predictions feed into shared auctions where delivery capacity and pacing controllers jointly determine which impressions are served~\cite{conitzer2022fppe,balseiro2024fieldguide,stram2024mystique}, and even a one-percentage-point shift from high-priority to lower-priority objective buckets can materially affect aggregate outcomes~\cite{wang2023pinterestbias,zheng2025adaptationbias}. No existing offline evaluation method provides visibility into these impression share shifts before online evaluation. We address this gap by proposing \emph{impression share prediction}: given a candidate ranking model and the current system state, predict the distribution of impressions across objective buckets it would produce in an online environment. To our knowledge, no existing work addresses this prediction task directly.}


\par{One challenge is that impression share prediction is inherently counterfactual. The candidate model has never served traffic in the live system: it has not influenced delivery capacity, pacing controllers have not adapted to it, and its model confidence has not been shaped by data it generated. The impression shares we observe reflect the current model's imprint on the system, not what a new model would produce. We frame this as a counterfactual prediction task and propose a statistical learning framework that uses the candidate's model confidence features alongside current system state to estimate the impression shares it would induce.}

\par{We validate this approach on data from multiple ranking model families, predicting impression shares observed during A/B tests from a candidate's earliest-online signals as a proxy for online evaluation inputs. Our goal is not to replace reward-based offline evaluation, but to complement it: impression share prediction surfaces allocation behavior that current offline evaluation does not reveal.}

\par{Our contributions are as follows:
\begin{enumerate}
\item We propose impression share prediction as an offline evaluation task that gives practitioners visibility into a candidate model's allocation behavior before A/B testing.
\item We propose a structural causal model that formalizes how model predictions, delivery capacity, and pacing jointly determine impression allocation, and show the counterfactual effect is identified from observational data.
\item We develop a statistical learning framework and show that for models already seen in the system, it reduces L1 prediction error by 49\% over a constant baseline.
\item We identify a systematic failure in the first hour after a held-out model's first appearance in the system, and introduce an encoder-conditioned architecture that recovers this gap, achieving 22\% L1 improvement precisely where the baseline fails.
\end{enumerate}
}

\section{Related Work}

\par{Standard offline evaluation of ranking models relies on predictive metrics computed over logged impressions: normalized entropy (NE), relative information gain (RIG), AUROC, and group-level AUROC (GAUC)~\cite{castells2022offline,hidasi2023flaws,facebook_practical_ctr,zhou2018din,wide_deep,esmm}. These measure a model's accuracy on data generated under a different ranking policy: they do not capture how the model would redistribute impressions across objective buckets in an online environment, or whether that redistribution would benefit or harm downstream utility. Impression share prediction is a complementary offline task that addresses this gap directly. A parallel line strengthens offline-metric fidelity through unbiased learning-to-rank~\cite{joachims2017unbiased,oosterhuis2022dr} and safe-deployment objectives~\cite{gupta2023safe}, but still targets per-impression accuracy on a fixed exposure, not how exposure redistributes once a candidate is serving.}

\par{A related line studies the auction and pacing mechanisms that govern impression allocation in ranking systems~\cite{conitzer2022fppe,balseiro2019repeated,balseiro2024fieldguide,stram2024mystique}, including interference bias when experiments share delivery capacity~\cite{liao2023ab,liao2025interference} - a structural feature we also exploit in our causal model. These characterize allocation as a system-level equilibrium property but do not predict the specific impression shares a new ranking model would induce.}

\par{Several works study offline-to-online gaps and the detection of impression or exposure shift~\cite{jeungen2021topk,jeunen2023positionbias}. \citet{zheng2025adaptationbias} formalize algorithm adaptation bias, where partial-rollout A/B tests misstate full-deployment impact. \citet{wang2023pinterestbias} show a retrieval model can improve aggregate offline metrics while substantially reallocating impressions across objectives. Off-policy evaluation methods~\cite{li2011replay,swaminathan2015crm_jmlr,dudik2011dr,jeunen2024deltaope}, including auction-simulation benchmarks~\cite{su2024auctionnet} and distributional or pessimistic estimators~\cite{wu2023distope,sakhi2024logsmooth}, estimate counterfactual reward from logged data but assume stationarity and policy overlap and target reward rather than allocation across objective buckets. These assumptions break down when models interact through shared delivery capacity and pacing. None of these predict the impression shares a candidate model would produce before online evaluation; that is the task we address.}

\section{Method}
\label{sec:method}

\subsection{Problem Formulation}
\label{sec:problemformulation}

\par{We consider a ranking system where impressions are allocated across $C$ objective buckets (e.g., click-, conversion-, view-optimized). Multiple ranking models operate simultaneously, each covering a subset of objective buckets. Models are evaluated through parallel A/B tests: each arm serves a different candidate model alongside shared incumbent models that collectively cover all $C$ objective buckets. Each campaign has a finite \emph{delivery capacity}, set externally at the campaign level and drawn down by a pacing system regardless of which arm serves the impression, so all arms draw from the same capacity pool. This shared pool is the structural coupling between arms that makes impression allocation a system-level outcome rather than a per-model property. We use historical A/B test data--observed hourly snapshots of impression shares across objective buckets, model confidence signals, and delivery capacity states--to train a statistical model that predicts the impression shares a new candidate model would produce before it enters the live system. Formally, given current system state and a candidate model $m$, we predict the normalized impression share vector $\hat{Y}^m = (\hat{Y}^m_1, \ldots, \hat{Y}^m_C) \in \Delta^{C-1}$, where $\Delta^{C-1} = \{\mathbf{v} \in \mathbb{R}^{C}_{\geq 0} : \sum_{c=1}^{C} v_c = 1\}$ is the $(C{-}1)$-dimensional probability simplex and $\hat{Y}^m_c$ is the fraction of impressions allocated to objective bucket $c$.}

\subsection{Causal Model}
\label{sec:causal}

We formalize the system via a structural causal model (SCM)~\citep{pearl2009causality} over three time-varying variables:

\begin{itemize}
  \item $A^m_t$: \textbf{Model state} (treatment) - the \emph{model-determined} (intrinsic) prediction score distribution and calibration of model $m$, fixed by its architecture and training rather than by the traffic it serves; this is the object the identification argument requires, and it is what makes the missing edge $D_t \not\to A^m_t$ hold.
  \item $Y^m_t$: \textbf{Impression share} (outcome) - fraction of impressions allocated to each of the $C$ objective buckets, $Y^m_t \in \Delta^{C-1}$.
  \item $D_t$: \textbf{Delivery capacity state} (shared covariate) - remaining delivery capacity across all objective buckets, shared across all A/B arms.
\end{itemize}

\begin{figure}[h]
\centering
\includegraphics[width=.45\textwidth]{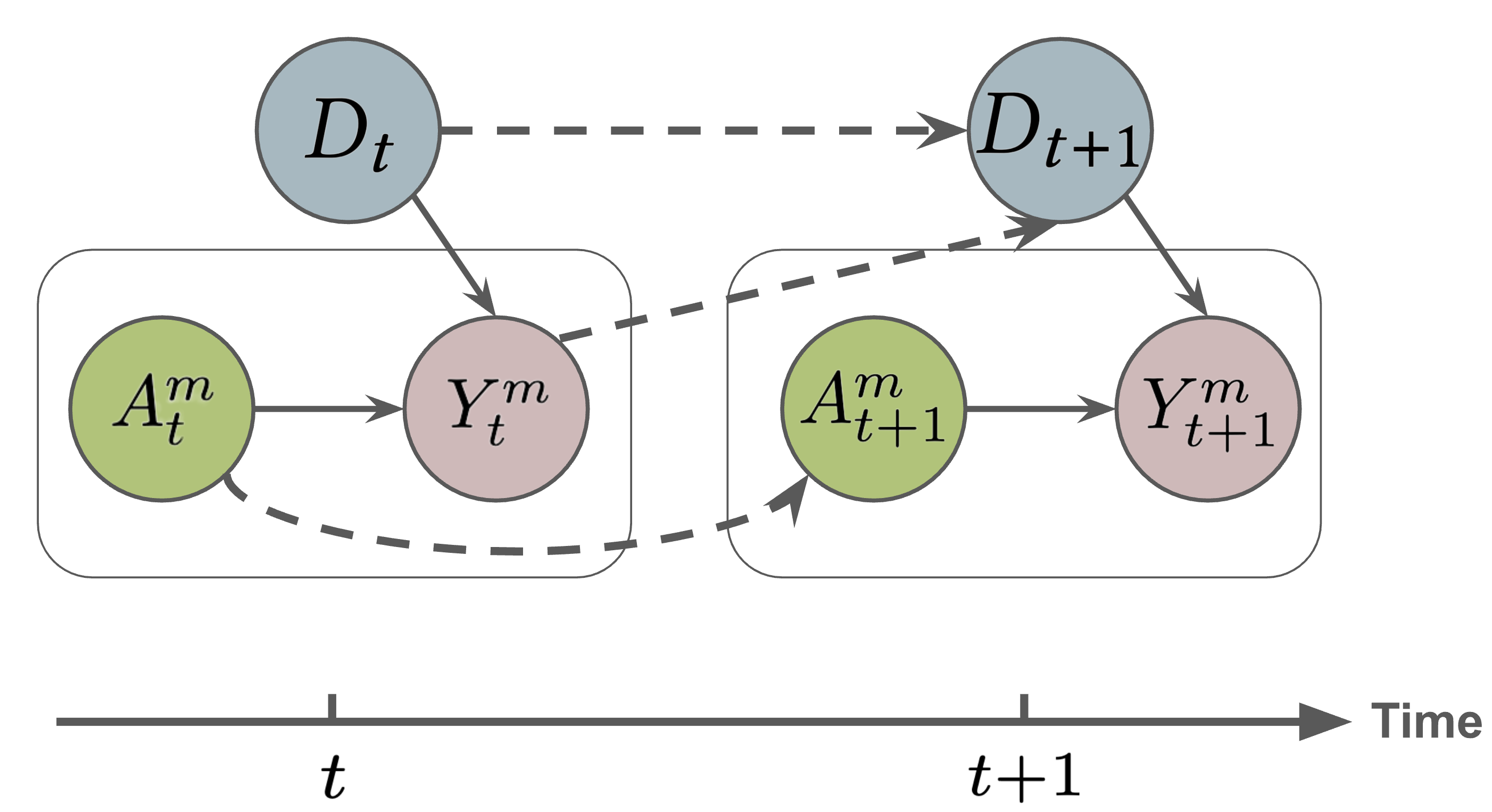}
\caption{Causal DAG for impression share prediction. $A^m_t$: model state (treatment); $Y^m_t$: impression share (outcome); $D_t$: shared delivery capacity state. Solid edges: within-period effects; dashed edges: cross-period dependencies.}
\label{fig:dag}
\end{figure}

Figure~\ref{fig:dag} shows the DAG. The edges encode: (1) $A^m_t \to Y^m_t$: model confidence determines auction winners, restricted to the model's covered objective buckets $V^m$; (2) $D_t \to Y^m_t$: capacity depletion and pacing multipliers constrain eligibility; (3) $Y^m_t \to D_{t+1}$: impressions served deplete shared delivery capacity; (4) $D_t \to D_{t+1}$: remaining capacity carries forward; (5) $A^m_t \to A^m_{t+1}$: model architecture is fixed, producing autocorrelated states.

One structural property is the absence of an edge $D_t \not\to A^m_t$: a model's properties are determined by its architecture and training data, not by system conditions. This eliminates all backdoor paths from treatment to outcome, so the causal effect of an intervention on the model state is identified from observational data: $P(Y^m_t \mid do(A^m_t = a), D_t) = P(Y^m_t \mid A^m_t = a, D_t)$. Unlike settings with time-varying confounding~\citep{melnychuk2022causal,shalit2017estimating}, no balanced representations, inverse propensity weighting, or domain-adversarial training are required.

\subsection{Predictive Framework}
\label{sec:predictive-framework}

\par{We frame impression share prediction as estimating $P(Y^m_T \mid A^m_0, D_0)$ from historical A/B data, using the variable definitions of Section~\ref{sec:causal}. Given a candidate model $m$, we observe its model state $A^m_0$ and the delivery capacity state $D_0$ at evaluation time $t=0$, and predict the impression share vector $Y^m_T \in \Delta^{C-1}$ aggregated over the forward window $[0, T]$ (we use $T = 24$h in our experiments). At this level of abstraction, $A^m_0$ is summarized by \textbf{model features} describing the candidate's prediction behavior, and $D_0$ is summarized by \textbf{delivery capacity features} describing the system at $t=0$; concrete feature definitions are deferred to Section~\ref{sec:setup}.}

\par{We distinguish two prediction regimes, post-entry and pre-entry. In the \emph{post-entry} setting the candidate has already been interacting with the live system, so $A^m_0$ and $D_0$ are both directly observable from logs at any time $t=0$, and we predict its share vector over the subsequent $T$ hours. In the \emph{pre-entry} setting the candidate has not yet served live traffic: it has not influenced delivery capacity through its consumption patterns, pacing controllers have not adapted to its behavior, and its confidence scores have not been shaped by data it generated~\citep{zheng2025adaptationbias}. The model state $A^m_0$ is model-determined -- ideally computed offline before online evaluation, and in practice estimated from the candidate's earliest online serving (Section~\ref{sec:setup}). For the delivery capacity input we use the \emph{current} system state $D_0$: we cannot know what the system will look like at the moment the candidate actually enters, so we ask what the candidate would produce under today's observed conditions:
\begin{equation}\label{eq:counterfactual-approx}
P\bigl(Y^{m}_T \mid \underbrace{A^{m}_0}_{\text{model-determined}},\; \underbrace{D_0}_{\text{current system state}}\bigr).
\end{equation}
Predicting $Y^m_T$ exactly from only $t=0$ inputs would otherwise require predicting the intermediate states $(A^m_t, D_t, Y^m_t)$ for $0 < t < T$, which compounds errors across steps. We instead learn predictors that map $(A^m_0, D_0)$ directly to $Y^m_T$, treating the intermediate dynamics implicitly through the historical training data. Since the true pre-entry outcome is unobservable, for empirical evaluation we use the candidate's \emph{first-hour} observed impression shares -- the window before the system has measurably adapted to it -- as a surrogate for the pre-entry target.}

\enlargethispage*{\baselineskip}
\section{Experiments}
\label{sec:production}

\subsection{Dataset and Setup}
\label{sec:setup}

\par{The system runs multiple concurrent A/B tests, each evaluating multiple candidate ranking models in parallel. We capture minute-level snapshots of auction statistics and impression shares across $C$ objective buckets from 150 candidate models spanning multiple model families, collected over roughly five weeks ($\sim$1.8M snapshots). We split the data temporally at a fixed cutoff into 23 training and 15 test days. Of the 103 models active in the test window, 20 are newly entered (never seen in training) and 83 were also present in the training period, letting us report performance separately on seen and unseen models. We evaluate under the two regimes introduced in Section~\ref{sec:predictive-framework} (post-entry and pre-entry). Predictions are compared against a \emph{constant baseline} that emits the training-set mean share vector for every test instance. We report two metrics: \textbf{L1 distance} $\mathrm{L1}(\hat{Y}^m_t, Y^m_t) = \sum_c |\hat{Y}^m_{t,c} - Y^m_{t,c}|$, the primary error metric (lower = better), and \textbf{Spearman rank correlation} between predicted and observed share vectors, a secondary metric (higher = better). Per-method scores are reported as relative improvement over the constant baseline.}

\par{\textbf{Feature instantiation.} The model state $A^m_0$ is realized as \textbf{model confidence features} (22-dim) -- a 20-bin log-scale histogram of the candidate's prediction scores, plus mean and variance -- together with a \textbf{coverage vector} $V^m \in \{0,1\}^C$ encoding which objective buckets the candidate is eligible to score. Ideally these scores would come from an offline evaluation pass before the candidate serves any traffic -- a full \emph{pre-entry offline evaluation}; in this work we instead estimate them from a rolling window over the candidate's earliest online serving, so our setup is \emph{early-entry forecasting}. For each objective bucket, the delivery capacity state $D_0$ is realized as \textbf{delivery capacity features} ($5 \times C$-dim): total capacity, consumed capacity, remaining capacity, pacing multiplier, and number of active campaigns.}

\subsection{Predictors}
\label{sec:predictors}

\par{We evaluate two predictors mapping $(A^m_0, D_0)$ to the forward share vector $\hat{Y}^m_T$. The first is a \emph{Random Forest} regressor that predicts each share component independently and then projects onto the simplex (clip negatives, renormalize). It treats the inputs as a single snapshot at $t=0$, with confidence features as the dominant signal and delivery capacity features contributing secondary predictive value. Two models with identical AUROC can produce very different score distributions and therefore very different impression shares.}

\par{The second is an \emph{encoder-conditioned} architecture that augments the snapshot view with a short look-back over recent capacity dynamics. A PatchTST-style encoder~\cite{nie2022time} ingests a 2-hour capacity history at minute-level resolution, divides it into 15-minute patches (8 tokens), and processes them through a 2-layer Transformer (4 attention heads, $d_\mathrm{model}=64$, GELU) with mean pooling to produce a system state vector $\mathbf{h}_t \in \mathbb{R}^{64}$. A conditional MLP head maps $(\mathbf{h}_t, A^m_0)$ to $\hat{Y}^m_T$ via two hidden layers followed by softmax. The encoder absorbs the dynamics over the immediate past, while the conditional head treats the model state as a separate intervention input -- enabling counterfactual queries (fix $\mathbf{h}_t$, swap the model state, read off predicted shares). Trained end-to-end with KL divergence loss and AdamW.}

\subsection{Evaluation on Seen Models}
\label{sec:onpolicy}

\par{We first evaluate on seen models: every test model also appears in training (the post-entry regime of Section~\ref{sec:predictive-framework}, where the model has been interacting with the system and we predict 24 hours into the future). This experiment serves two purposes: (i) establish how predictable the task is when the model is in-distribution, and (ii) ablate the contribution of each feature group. We compare three Random Forest configurations differing in feature set against the constant baseline; Table~\ref{tab:onpolicy} reports relative improvements over the baseline.}

\par{Confidence and coverage features alone improve L1 by 42.9\%, accounting for the majority of the gain. Delivery capacity features alone improve L1 by a modest 25.3\%. The full feature set improves L1 by 48.6\%. The candidate model's score distribution - not the system state - is therefore the primary predictor of impression allocation when the model has been seen during training.}

\par{Spearman rank correlation is near-saturated across all configurations: the rank ordering of objective buckets by impression share is largely stable across model versions, since system structure (delivery capacity, campaign mix) determines which buckets are higher priority. L1 is the more consequential metric, since small share changes shift downstream utility (buckets differ in conversion rate).}

\begin{table}[!htb]
\centering
\small
\begin{tabular}{@{}lcc@{}}
\toprule
\textbf{Feature Set} & \textbf{L1 imp.} & \textbf{Spearman imp.} \\
\midrule
Constant Baseline & 0 & 0 \\
Capacity Only & $+$25.3\% & $+$0.17\% \\
Confidence + Coverage & $+$42.9\% & $+$0.64\% \\
Capacity + Confidence + Coverage & $+$48.6\% & $+$0.66\% \\
\bottomrule
\end{tabular}
\caption{Seen-models evaluation: relative improvement over the constant baseline (24h evaluation window). Higher = better. L1 is the primary metric.}
\label{tab:onpolicy}
\end{table}

\subsection{Evaluation on Held-Out Models}
\label{sec:heldout}

\par{This experiment has two ingredients. First, the test models are \emph{held out}: they never appear in training, so the predictor has no in-distribution signal for them. Second, because we use online data, the input we feed -- the model state $A^m_0$ and especially the system state $D_0$ -- itself depends on how long the model has been interacting with the live system. We therefore further break down evaluation by \emph{time since the model first appeared}. The same held-out model yields very different inputs depending on whether $D_0$ is captured one hour after entry or five days after: longer interaction means $D_0$ has progressively absorbed the model's own footprint on capacity consumption and pacing, so even though the model is unseen by the predictor, the input has become much more predictable. The true offline-to-online gap lives in the \textbf{first hour}: the features extracted there represent the offline setup -- the model has not yet measurably impacted the system, and $D_0$ still reflects the prior regime.}

\par{Table~\ref{tab:heldout} reports L1 improvement over the constant baseline for the two predictors -- the snapshot Random Forest and the encoder-conditioned PatchTST Transformer (Section~\ref{sec:predictors}) -- binned by time since first appearance. One distinction is that the Transformer's encoder simulates a short rollout over the most recent 2 hours of auction dynamics, whereas the RF uses only the instantaneous snapshot at $t=0$ -- no rollout, neither backward over the recent past nor forward over the 24-hour prediction window. At 0--1h the RF is $-$20.5\% (worse than the constant baseline) because $D_0$ still encodes the prior model's equilibrium and the rolling confidence window is only partially filled; it then gradually improves, recovering to $+$11.5\% by 1--2h and converging to $+$37.9\% by 7d$+$. The Transformer's 2-hour capacity rollout closes the first-hour gap directly: at 0--1h it achieves $+$22.1\% -- a \textbf{42.6 percentage-point swing} against the RF -- and it leads through 12h. Beyond 2 days the RF catches up and slightly leads: by then the system has fully adapted to the new model and instantaneous features suffice.}

\begin{table}[!htb]
\centering
\small
\begin{tabular}{@{}lrr@{}}
\toprule
\textbf{Time since entry} & \textbf{RF \%imp} & \textbf{Transformer \%imp} \\
\midrule
0--1h   & $-$20.5\% & $+$22.1\% \\
1--2h   & $+$11.5\% & $+$33.2\% \\
2--4h   & $+$27.2\% & $+$27.5\% \\
4--8h   & $+$29.2\% & $+$37.1\% \\
8--12h  & $+$29.6\% & $+$37.1\% \\
12--24h & $+$24.8\% & $+$34.1\% \\
1--2d   & $+$30.8\% & $+$32.3\% \\
2--4d   & $+$26.2\% & $+$25.7\% \\
4--7d   & $+$36.4\% & $+$34.6\% \\
7d$+$   & $+$37.9\% & $+$33.8\% \\
\bottomrule
\end{tabular}
\caption{L1 improvement over the constant baseline on held-out (unseen) models, by time since first appearance. The 0--1h row corresponds to the pre-entry surrogate. Positive = better than baseline. Rank correlation is saturated (Spearman $>0.98$) across all bins and is omitted.}
\label{tab:heldout}
\end{table}

\section{Conclusion}

\par{We introduced impression share prediction as an offline evaluation task for ranking systems, complementing reward-based metrics by surfacing how a candidate model would redistribute impressions across objective buckets before online evaluation. We formalized the task via a structural causal model that shows the effect is identified from observational data, requiring neither inverse propensity weighting nor balanced representations, and framed prediction as estimating $P(Y^m_T \mid A^m_0, D_0)$ over a 24-hour forward window. Two studies on data examine (i) feature importance under in-distribution conditions, and (ii) prediction accuracy on held-out models, broken down by how long the model has been interacting with the system. The first hour is where the offline-to-online gap actually lives: a snapshot Random Forest falls below the constant baseline ($-$20.5\% L1), while a PatchTST encoder that simulates a 2-hour rollout over recent auction dynamics recovers $+$22\% L1 improvement precisely in this regime.}

\par{Two directions follow. First, extending the predictor to a full rollout from $t=0$ to $T$ requires a Bayesian or sequential model that propagates uncertainty across $(A^m_t, D_t, Y^m_t)$ and handles compounding error -- the present work avoids this by direct prediction. Second, the most accurate setup would compute the candidate's confidence features from offline evaluation; in this work we use the rolling online-window confidence as a stand-in, and any discrepancy between offline-evaluation confidence and online rolling confidence could impact results in the pre-entry regime.}

\bibliographystyle{ACM-Reference-Format}
\bibliography{camera_ready_submission}

\end{document}